\documentclass{aa}     
\usepackage{epsf}
\usepackage{amssymb}
\usepackage{amsmath}
\usepackage{amsxtra}
\usepackage{times}
\usepackage{graphicx}
\usepackage{rotating}
 
\def\aa{A\&A}

\begin{document}

   \title{Limits on the Transverse Velocity of the 
   Lensing Galaxy in Q2237$+$0305 from the Lack of Strong 
   Microlensing Variability
   }

   \author{Rodrigo Gil-Merino$^{1}$, 
	Joachim Wambsganss$^2$, 
	Luis J. Goicoechea$^1$,
	Geraint F. Lewis$^3$
   	}
 
   \institute{$^1$  Universidad de Cantabria, Departamento de F\'{\i}sica Moderna,
		  Avda. de Los Castros s/n, E-39005 Santander;
		  \email{gilmerinor, goicol@unican.es} \\
	$^2$ Astronomisches Rechen-Institut (ARI) and Universit\"at Heidelberg, 
		M\"onschhofstr. 12-14, 69120 Heidelberg, Germany;
		  \email{jkw@ari.uni-heidelberg.de} \\		  
	$^3$  University of Sydney, Institute of Astronomy, School of Physics                                         
   		 A28, NSW 2006, Australia;
		  \email{gfl@physics.usyd.edu.au}  \\
		  }
		  
   \offprints{R. Gil-Merino}
 
\date{Received date; accepted date}
 
\titlerunning{The transverse velocity of the Q2237$+$0305 lens}
\authorrunning{Gil-Merino et al.}
 
\abstract{We present a method for the determination of upper
limits on the transverse velocity of the lensing galaxy in the quadruple 
quasar system Q2237$+$0305, based on the lack of strong microlensing 
signatures in the quasar lightcurves.
The limits we derive here are 
based on four months of high quality monitoring data, 
by comparing the low amplitudes of the lightcurves of the 
four components  with extensive numerical simulations. We make use 
of the absence of strong variability of the components 
(especially components B and D) to infer that a ``flat'' time
interval  of such a 
length is only compatible with an effective transverse velocity 
of the lensing galaxy of 
$    {v_{\rm bulk}}~\leq~630$ km/s for typical 
microlenses masses of $M_{\rm \mu lens}=0.1M_{\odot}$ (or 
$    {v_{\rm bulk}}~\leq~2160$ km/s for 
$M_{\rm \mu lens}=1.0M_{\odot}$) 
at the 90\% confidence level. This method may be applicable in the 
future to other systems.
\keywords{
   Gravitational lensing
-- Microlensing
-- Quasars: Q2237$+$0305
-- General: data simulations
}
}


\maketitle

\section{Introduction}

Measurements of the  peculiar motions of  galaxies can provide
strong constraints on the nature  of dark matter and the formation and
evolution  of   structure  in   the  Universe.    
Determining such `departures from the  Hubble flow', utilizing
standard distance indicators as the Tully-Fisher relation for spirals
and the $D_n-\sigma$ method for ellipticals (e.g., Peebles 1993), have
proved  to be quite  difficult.  While  these measures  provide radial
peculiar motions,  transverse peculiar motions are also  required to fully
constrain  cosmological  models. However,  the   determination  of
transverse velocities is an extremely difficult task, generally
beyond  the reach of  current technology.   Recently, Peebles  et al.
(2001) suggested  the  use of  the  space  missions  SIM and  GAIA  to
estimate the transverse displacements of nearby galaxies.  Roukema and
Bajtlik  (1999) claimed  that  transverse galaxy  velocities could  be
inferred from  multiple topological images, under  the hypothesis that
the `size' of the Universe  is smaller than the apparently `observable
sphere'.  In  spite of these efforts, our knowledge of 
transverse motions of galaxies is currently very limited.

Dekel et al. (1990) showed that the local galaxy velocity field can be
reconstructed assuming that this  field is irrotational, and thus, the
measurement of  the transverse velocities  could be used to  test this
assumption. In  fact, the determination  of transverse motions  would
be very useful  to  discuss the  quality of the whole 
reconstruction.  
From another point  of view, the reconstruction methods  
are powerful tools to estimate  galactic transverse motions.   

Grieger,  Kayser  and   Refsdal  (1986)  had   suggested  using
gravitational  microlensing  of   distant  quasars  to  determine  the
transverse  velocity  of 
the  lensing  galaxy  via  the  detection  of  a
`microlens  parallax' as  the  quasar is  magnified  during a  caustic
crossing (see  also Gould 1995).  For the determination of  this 
parallax, however, ground-based monitoring is not sufficient, 
it requires parallel measurements from a satellite located at 
several AU in addition.

The gravitational lens Q2237$+$0305 consists of four images of
a  quasar at a redshift of $z_q=1.695$, 
lensed  by a  low  redshift ($z_g=0.039$)  spiral
galaxy  (Huchra   et  a.   1985).    Photometric  monitoring  revealed
uncorrelated  variability between the  various images,  interpreted as
being  due to  gravitational microlensing  (Irwin et  al.  1989). This
interpretation was confirmed with dedicated monitoring programs (e.g.,
{\O}stensen  et  al.  1996;  Wo{\'z}niak et  al.  2000a,b;  
Alcalde  et al. 2002). 
Q2237$+$0305 is the best studied quasar microlensing system.
With ten  years of monitoring  data, Wyithe  et al.
(1999) used  the  derivatives  of  the  observed  microlensing
lightcurves to  put limits  on the lens  galaxy tranverse  velocity of
Q2237+0305. Very recently, Kochanek (2004) developed a method for 
analysing microlensing lightcurves based on a $\chi^2$ statistics which 
includes the transverse velocity of the source as an output parameter.

Here we present another  method to determine upper limits
on the transverse velocity of  a lensing galaxy 
in a multiple quasar system. We apply it to 
Q2237$+$0305, based on a comparison  between 
four months of high  quality photometric monitoring   of  
the   four  quasar   images  and   intense  numerical
simulations.  
The details  of the  microlensing model and simulations 
are  discussed in Section  2.  
In  Section 3  we  briefly present and review
the lens  monitoring results and
outline our method  to obtain limits on the  transverse velocity.  
The results  of this  approach \;-- 
the constraints on the transverse velocity  of the lensing galaxy
--\;  are  presented in  Section 4 and 
discussed in Section 5. We include our conclusions in Section 6.

\section{Microlensing background}

\subsection{Lens models of Q2237+0305}

Several  approaches have been employed in  modeling the observed
image configuration  in Q2237+0305  (Schneider et al. 1988, Wambsganss 
and Paczy\'nski 1994, Chae et al. 1998, Schmidt et al. 1998). 
These models  provide the
parameters relevant to microlensing studies: the surface mass density,
$\Sigma$, and the  shear, $\gamma$, at the positions  of the different
images. The  former represents the  mass distribution along  the light
paths projected into  the lens plane, while the  latter represents the
anisotropic  contribution of  the  matter outside  the  beams. We  can
normalize  the surface  mass density  with the  critical  surface mass
density (see Schneider et al. 1992 for more details),
\begin{equation}
\Sigma_{crit}=\frac{c^2}{4 \pi G}\frac{D_s}{D_d D_{ds}}
\label{sigcrit}
\end{equation}
where  $D_s$,  $D_d$ and  $D_{ds}$  are  the  angular  diameter
distances between  observer and  source, observer  and 
deflector and between deflector and source, respectively, $c$
is the  velocity of light and  $G$ is the  gravitational constant. The
resulting    normalized   surface    mass    density   (also    called
\emph{convergence}   or   \emph{optical   depth})  is   expressed   as
$\kappa=\Sigma/\Sigma_{crit}$.

We use here two different sets of values for $\kappa$ and $\gamma$ for
the   four  components   (Tab.~\ref{table1}),  corresponding   to  the
Schneider et  al. (1988)  and the Schmidt  et al. (1998)  lens models,
respectively.  
We will demonstrate  using these two sets that slightly
different values  for the two  local lensing parameters do  not change
the results, and hence that some scatter in $\kappa$ and $\gamma$
of the images does not affect the conclusions.

\begin{table}[htb]
\centering
\begin{tabular}{cccccc}
\hline\noalign{\smallskip}
       & \multicolumn{2}{c}{Schneider et al. (1988)} & & 
	 \multicolumn{2}{c}{Schmidt et al. (1998)}\\
 Image & $\kappa$ & $\gamma$ & & $\kappa$ & $\gamma$\\
\noalign{\smallskip}\hline\noalign{\smallskip}
 A & 0.36 & 0.44 & & 0.36 & 0.40 \\
 B & 0.45 & 0.28 & & 0.36 & 0.42 \\
 C & 0.88 & 0.55 & & 0.69 & 0.71 \\
 D & 0.61 & 0.66 & & 0.59 & 0.61 \\
\noalign{\smallskip}\hline
\end{tabular}
\caption{Two different  sets of values  for the surface  mass density,
$\kappa$, and  the shear,  $\gamma$, of the  four images are  used, in
order  to study  the dependence  of the  result on  the lens  model (see
References for details).}
\label{table1}
\end{table}

\subsection{Simulations}

We  use the  ray shooting  technique  (see Wambsganss  1990, 1999)  to
produce  the 2-dimensional magnification  maps for  each  of the
gravitationally lensed images.  
All the  mass is assumed to be in 
compact  objects,  with  no  smoothly
distributed  matter.  This should be a good approximation
for an old stellar population in the central part of the 
lens galaxy and a small amount of smooth matter would not introduce a
significant difference in our results. All  of the  microlensing  
objects  are assumed to  
have a mass of  $M_{\rm \mu lens}$ and  are distributed randomly
over the lens plane.  Taking  into account the effect of the shear
and  the combined deflection  of  all microlenses,  light rays  are
traced  backwards
from the  observer to  the source.  This  results  in a
non-uniform  density of rays  distributed over  the source  plane. The
density of rays  at a point is proportional to the  microlensing
magnification of a source at that position; hence
the result of  the rayshooting technique is a  map of the microlensing
magnification  as a  function of  position in  the source  plane.  The
relevant scale  factor, the Einstein  radius in the source  plane, is
defined as
\begin{equation}
r_E=\left(\frac{4GM_{\rm \mu lens}}{c^2}\frac{D_s
D_{ds}}{D_d}\right)^{1/2} .
\label{re}
\end{equation}
Finally, the magnification pattern is convolved with a particular
source  profile for the quasar.   
Linear  trajectories  across  this  convolved  map,
therefore, result in microlensing  light curves (see also Schmidt \& 
Wambsganss 1998).

In general, the details of a quasar microlensing light curve depend on
several unknown  parameters: the masses and positions  
of the microlenses
and the size, profile and effective transverse velocity of the source.
For  this   reason,  the  comparison  of   the  simulated  microlensing
lightcurves  to the  observed ones  must be done in  a statistical sense.

\section{The Method}

\subsection{The idea in a nutshell}

\begin{figure}[hbtp]
 \centering \epsfxsize=6.00 cm \rotatebox{0}{\epsffile{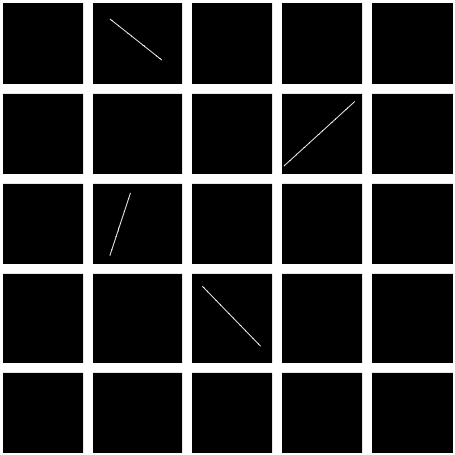}}
 \caption[]{Idealized  magnification pattern to  illustrate the
 idea  of  the  method:  
 Black  areas are low  magnification  zones,
 the regular grid of 
 thick white lines  represents the caustics (high  magnification 
 areas)  and the
 thin white lines are example
 tracks due to the relative motion between source, lens
 and observer, 
 which all would result in {\em flat} lightcurves.}  \label{ideal}
\end{figure}

Before going  into details of  the method we  use, we present  a very
simple hypothetical scenario to better illustrate the procedure.
Generally, microlensing  magnification maps  possess significant
structure, in particular they consist of an intricate web 
of very high magnification regions, the caustics. 
The density and the length of the caustics vary with the
values of surface mass density $\kappa$ and shear $\gamma$. However,
for a given pair of parameters 
$\kappa$ and $\gamma$, there is something like a {\em typical}
distance between adjacent caustics (Witt et al. 1993), 
though with quite a large dispersion. 
For illustration purposes, we assume now 
that we have  a magnification pattern
with caustics  that are equally spaced  horizontal and vertical
lines (see Fig.~\ref{ideal}).  
Though this is far from being a realistic magnification
pattern,   
its  simplicity allows us to explain the
relation between fluctuations in  the microlensing lightcurves and the
velocity of  the source in simple terms.  
The pattern shows schematically  the typical
low (dark)  and high  (white) magnification areas,  respectively.  
The length and width of the  low  
magnification areas  is exactly one unit   length, $l_{\rm unit}$.
If  we compute the  magnification along 
a  linear track {\em inside} one of  these regions,
the resultant  lightcurve will be flat.  
However, there is a maximum length for such flat lightcurves:
there cannot be any flat lightcurves with lengths larger than
$l_{\rm max}  = \sqrt{2}~l_{\rm unit}$.  
Now  suppose that this  magnification map
corresponds  to a certain  hypothetical gravitationally  lensed system
and we  have a flat observed microlensing  lightcurve corresponding to
an observing period of $t_{\rm obs}$. 
Then we can calculate an upper limit
for the velocity of the source: 
$V_{\rm max} = \sqrt{2} l_{\rm unit} /t_{\rm obs}$.
Of course, the actual velocity could be (much) smaller
than that;  if only ``flat'' microlensing lightcurves
existed, it could be as small as zero (measured caustic
crossings will provide lower limits on the velocity).
As  stated above, true microlensing magnification  maps are much
more complex than the idealized case presented in Figure~\ref{ideal}.
Nevertheless, we  can  determine an upper limit 
on the track lengths for a realistic
magnification pattern in a statistical sense, 
by just replacing
the {\em fixed} distance between the idealised
caustics by the real distribution of caustic distances, 
and the perfectly flat parts for the regions between the caustics by
the relatively flat real lightcurves. 
This way we can get an 
upper limit $l_{\rm upper}$ on the track  lengths (in $r_E$)
that is consistent with the observed variability.  
Since we know  the duration  of the  observing
period $t_{\rm obs}$ from the actual monitoring campaign,   
it is straightforward to obtain  the upper limit on the
transverse velocity   for assumed  
values of the lens mass $M_{\rm \mu  lens}$ and the source size/profile:
$V_{\rm upper} = l_{\rm upper}/t_{\rm obs}$. 

Although the effective transverse velocity has contributions from
all three components source, lens, and observer as shown below, 
for the system  Q2237+0305, the effective
transverse velocity is dominated by the effective 
transverse  velocity of the lensing galaxy (Kayser et al. 1986).

\subsection{Monitoring Observations of Q2237+0305 to be compared with}

This  study employs  the results  of the  GLITP  (Gravitational Lenses
International Time  Project) collaboration which  monitored Q2237+0305
from October  1st, 1999  to February  3rd, 2000, using  the 2.56 m Nordic
Optical  Telescope (NOT)  at El  Roque de  los  Muchachos Observatory,
Canary  Islands, Spain  
(see Alcalde  et  al. 2002  for data  reduction details and results).

\begin{figure}[hbtp]
 \centering
 \epsfxsize=8.5 cm
 \rotatebox{0}{\epsffile{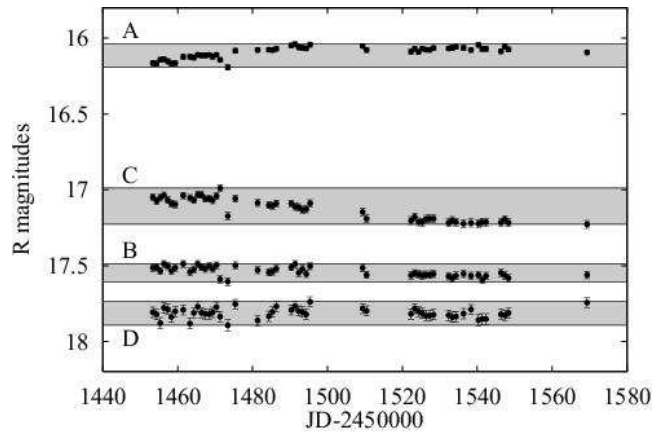}} 
 \caption[]{The  R  band  photometry  of  Q2237+0305  from  the  GLITP
 collaboration. The  observing period was  from October 1st,  1999 (JD
 2459452) to February  3th, 2000 (JD 2459577) with  the Nordic Optical
 Telescope   at  Canary   Islands,  Spain   (details  in   Alcalde  et
 al. 2002). The components are labeled from A to D (Yee 1988). The 
 bands indicate the amplitude of each component and are defined by the 
 maximum and the minimum magnitude in each lightcurve. The 
 widths of these bands are $\Delta m_A = 0.154$ mag, 
 $\Delta m_B = 0.116$ mag, 
 $\Delta m_C = 0.238$ mag and $\Delta m_D = 0.155$ mag.}
 \label{Rband} 
\end{figure}

The   R   band  photometry  used       here   is  shown   in
Fig.~\ref{Rband}. It  is clear that whereas components  A and C
show a small but significant  variability (see Shalyapin et al.  2002
and Goicoechea et al. 2003 for the interpretation of the variation in the 
component A as the peak of a high-magnification event),  
images B  and D  remain  relative flat,  showing no  signs of
strong microlensing during the monitoring period.  As the
expected  time  delays  between  the images  are  short  
($\le 1$~day, see Schneider et al. 1988, 
Wambsganss \& Paczy\'nski 1994),
intrinsic fluctuations  would show up in all  4 images almost
simultaneously.
Keeping in mind the idea expressed in the previous subsection, we used
the  relative flatness of  all four components of Q2237+0305
to  statistically  infer upper
limits  on   the  length  of   linear  tracks  in   the  corresponding
magnification patterns.

For a given component,   we define
the largest fluctuation in the lightcurve by the
difference between the 
maximum and the minimum magnitudes. 
Thus $\Delta m_X=|m_{X, max}-m_{X, min}|$, where X denotes 
component A, B, C or D. 
For the simulated microlensing lightcurves the condition to be
fulfilled then is: 
$\Delta m_X (simul) \le \Delta m_X$, where $\Delta m_X (simul)$ 
is the difference between the maximum and the minimum 
in the simulated lightcurve (again X stands
for the four components A, B, C or D). In this way, for each set of 
simulations, if at least one component shows fluctuations larger than 
observed, then the set is rejected.
For component A we obtained $\Delta m_A=0.154$ mag; for component B,
$\Delta m_B=0.116$ mag; for component C, $\Delta m_C=0.238$ mag
and  for  component D, $\Delta m_D = 0.155$ mag (see Fig.~\ref{Rband}).

\begin{figure}[hbtp]
 \centering
 \includegraphics[bbllx=172,bblly=209,bburx=440,bbury=583,width=5.0cm,
                  angle=-90,clip=true]{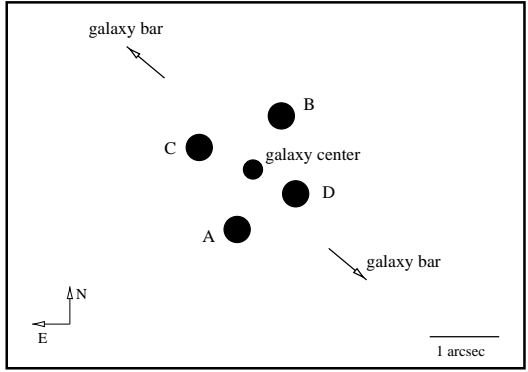}
 \caption[]{Relative positions of the quasar images, galaxy center and
 galaxy bar. The direction of motion relative to the external shear is
 not independent between the images because of the cross-like
 configurations. Thus, this is orthogonal between images A$\perp$C and
 B$\perp$D, whereas the direction for image A is parallel
 to the one in image B, and similarly for images C and D.
 }
 \label{config} 
\end{figure}
 
It is important to notice that if erronous
outliers would be present in the photometry, 
the result would be an overestimate:
the largest actual fluctuations would be 
lower than the obtained result. This would 
result in a larger upper limit on the transverse
velocity, hence the approach we use is conservative.
However, looking
at Figure 2, this seems an unlikely possibility.

\subsection{Microlensing Simulations}

We computed magnification patterns for the four quasar images,
using the Schmidt  et al. (1998) model for the  values of $\kappa$ and
$\gamma$  (cf. Table 1).
We assumed all compact  
objects have the same  mass, $M_{\rm \mu  lens}$ (realistic mass functions 
act roughly as their respective mass averages, so
we take $M_{\rm \mu  lens}=0.1M_{\odot}$ and $M_{\rm \mu  lens}=1M_{\odot}$ 
as extreme cases; see i.e. Lewis \& Irwin 1995).
The physical side
lengths  of  these  maps  were  $15~r_E$  covered  by  $4500$  pixels,
resulting in a spatial resolution  of 300 pixels per Einstein radius
$r_E$. For each component we did a number of different random realizations 
of maps of this size, to be sure that they were statistically representatives. 
The effect of the finite source size is 
included by convolving the magnification patterns with
a  certain source  profile. We  adopted  a Gaussian  
surface  brightness  profile for  the
quasar  with  three  different values  of the width 
$\sigma_Q$= $0.003$ $r_E$, $0.01$ $r_E$ and $0.05$ $r_E$. 
This corresponds to physical effective radii from 
$2\times10^{14}$ cm to $3\times10^{15}$ cm for 
$M_{\rm \mu  lens}=0.1M_{\odot}$, and a factor of $\sqrt{10}$ larger
for lens masses of $M_{\rm \mu  lens}=1.0M_{\odot}$. Much 
larger source sizes are excluded by the large amplitude
fluctuations in this system observed by Wo\'zniak et al. (2000a,b),
as was shown by Yonehara (2001) and Wyithe, Webster \& Turner (2000a).
Therefore we can restrict our analysis to small source sizes. In addition, 
we remark that the important contribution of the source in our simulations comes 
from its scale and more realistic models for the source are considered second order
effects. This can be better understood if we think that                                                             
'fine structure' of these realistic models will produce refined                                                                  
fluctuations in the simulated lightcurves.  But this fluctuations are                                                            
originated close to caustics and the method is based on the selection of                                                            
realizations  with low fluctuations, where this 'fine structure' does not exist.


\begin{figure}[hbtp]
 \centering 
  \includegraphics[bbllx=19,bblly=19,bburx=575,bbury=575,width=8.0cm, 
                  angle=-90,clip=true]{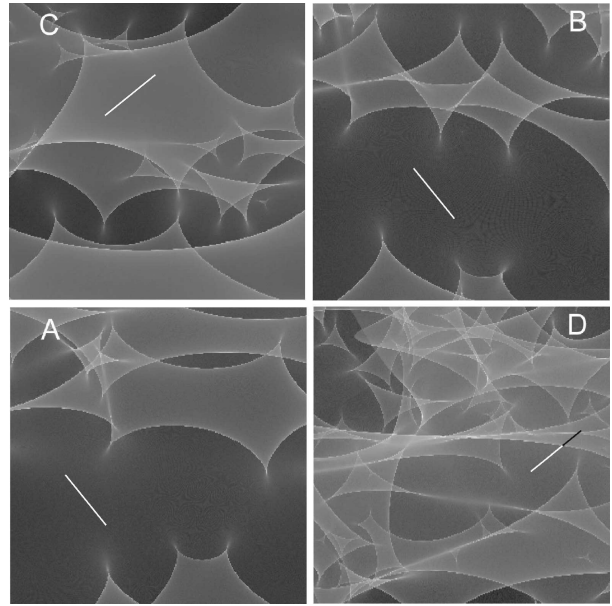}													
 \caption[]{A  part of  the total magnification  pattern  
 for each component, labeled and ordered as in Fig.~\ref{config}.  
 The  length  of the white part of the track is determined in such a 
 way as to fulfill the criterion 
 $\Delta m_X (simul) \le \Delta m_X$, where $X$ denotes the component A,
 B, C or D. The black part of the track in component D introduces 
 variability higher than the observed one, making this set not compatible 
 with the observations.}
 \label{magpat} 
\end{figure}

In order to statistically infer an upper limit on the permitted length
of the linear tracks across the magnification maps,
we do the following (for a given source size):
we first fix the length of the track and randomly select a starting 
pixel in the magnification  pattern of  one of the images, say component A.
Then we select a random  direction for which  the magnification along a 
linear  track is going to be computed. 
As the next step, 
random starting points in the magnification patterns
of the  other  images are selected.  However, this time
the direction is not arbitrary: The direction of motion
in all the images relative to the external shear is fixed, 
the displacements of  the  source in  the  
magnification  maps B, C and  D  are no  longer independent. 
In fact, because  of  the cross-like geometrical configuration  
of the system, A and B are parallel to each other, as
are C and D. These two pairs are orthogonal  relative to
each other: 
A$\perp$C and B$\perp$D (cf. Fig.~\ref{config} motivated by 
Fig.~1 in Witt and Mao, 1994). 
Thus, once  the direction in the magnification pattern
A is  selected, the ones in the magnification  patterns B, C, 
and D are determined as well. 
So in this way we construct simultanously sets of
test lightcurves
for the four quasar images along linear tracks 
for a given track length.
This is illustrated in 
Fig.~\ref{magpat}, where a small part of each  of  these
magnification patterns is shown (side  length is $L \approx 4~r_E$).
White color indicates high magnification 
while black means low magnification.   
The linear tracks drawn inside the maps in Fig.~\ref{magpat}
illustrate the calculation procedure.

When  $\Delta m_X (simul)$  
-- the amplitudes between maximum and
minimum of the simulated lightcurves for images A, 
B, C, and D
-- are larger than  $\Delta m_X$, 
then this particular set of lightcurves is marked as 
not compatible with the observations. 
In the example shown in Fig.~\ref{magpat},
the parts of the four tracks with 
amplitudes lower than the observed ones corresponds to the white 
part.

This procedure is performed for 50000 sets of
tracks/lightcurves per given track length, and
repeated it for different lengths, 
ranging from $0.01~r_E$ to $1~r_E$  
in steps of $0.003~r_E$
(which numerically corresponds to lengths of $3$ to $300$ pixels
in steps of one pixel). 
This corresponds to a total of $1.5\times10^7$ simulated tracks.
For each considered track length, we determine
then the fraction of tracks not  
compatible with the observations, i.e., 
the probability distribution.


From this distribution we can now derive $l_{\rm upper}$ from
 the probability $P(l \leq l_{\rm upper})$ =  90\%, 
i.e., the 90 per cent upper limit on the allowed
path  lengths (or similarly, the 68\% or 95\% levels).
The whole  procedure   was  repeated for  magnification
patterns constructed with the $\kappa$ and $\gamma$ values
of the Schneider et  al. (1988) model, see Table 1.
The results (presented in the next section) were indistinguishable.

\section{Results}
\begin{figure}[hbtp]
\centering
\includegraphics[bbllx=0,bblly=0,bburx=721,bbury=505,width=8.5cm,
                angle=0,clip=true]{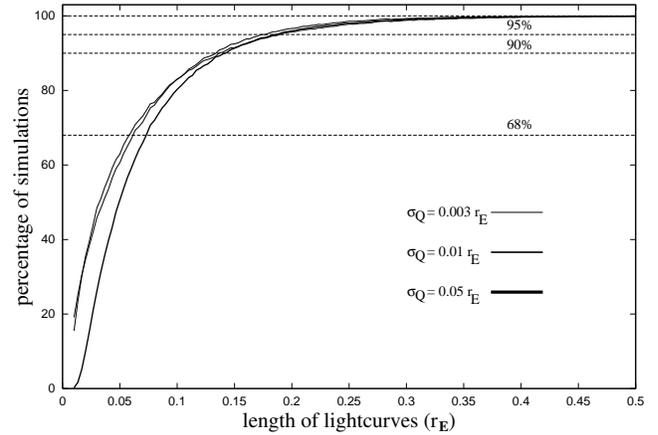} 
\caption[]{Probability distribution for the fraction of light curves
with a given length which produce larger fluctuations than what is observed.
We consider three different
source  sizes:  $\sigma_Q  = 0.003~r_E$ (thin line),
$\sigma_Q = 0.01~r_E$ (medium line), and
$\sigma_Q = 0.05~r_E$ (thick line). Horizontal dotted lines 
indicate the different cuts in percentages,
corresponding to the values listed in Table~\ref{tabresult}.}
\label{cumpro} 
\end{figure}

The resulting probability distributions for the
fraction of lightcurves which showed more fluctuations than
the observed ones are shown in Fig.~\ref{cumpro} for
each track length in units of Einstein radii.
The three curves represent the three different 
source  sizes we considered:   
$\sigma_Q=0.003~r_E$ (thin line),  
$\sigma_Q=0.01~r_E$  (medium line) and  
$\sigma_Q=0.05~r_E$ (thick line).

From each distribution we can determine the upper limit
on the length of the tracks consistent
with the variability of the  observed lightcurves defined by the bands
described   before.   
The values are given in Table~\ref{tabresult} for three
different ``confidence levels'' of 68\%, 90\% and 95\%, for
the three source sizes used. 
We estimate the uncertainty in $l_{upper}$(90\%)   
from the $\sqrt{N}$ statistics, 
where $N$ is the number of simulations at the 
90\% point: The error is $\pm 0.02~r_E$. 
The results are almost independent of the source size
(within the considered range),
as is apparent from Figure~\ref{cumpro} as well.
This is not too surprising, because in a ``flat'' part of the
magnification pattern, where most of our simulations will take place, 
a large source is (de-)magnified  by
the same amount as a small source (in the considered range of source 
sizes, the length of the tracks is always much larger than the size of 
the source). 
Only when a caustic is approached, this changes.
%
%
%

\begin{table}[htb]
\centering
\begin{tabular}{cccccc}
\hline\noalign{\smallskip}
 C.L. & \multicolumn{3}{c}{source sizes}\\
      & $\sigma_Q=0.003~r_E$ & $\sigma_Q=0.01~r_E$ & $\sigma_Q=0.05~r_E$\\
\noalign{\smallskip}\hline\noalign{\smallskip}
 68\% & 0.06~r$_E$ & 0.06~r$_E$ & 0.07~r$_E$ \\
 90\% & 0.13~r$_E$ & 0.13~r$_E$ & 0.13~r$_E$ \\
 95\% & 0.17~r$_E$ & 0.18~r$_E$ & 0.18~r$_E$ \\
\noalign{\smallskip}\hline
\end{tabular}
\caption{Results for the length of the simulated tracks 
$l_{\mathrm upper}$ as a function of source
size for different confidence levels; at the 90\% 
the uncertainty is $\pm 0.02~r_E$.}
\label{tabresult}
\end{table}

These  track lengths 
can be converted into  physical quantities by using  
Eq.~\ref{re} and  a given value  for the  mass of  the microlenses,  
$M_{\rm \mu lens}$. 
As  we are
observing the inner part of the lens galaxy with presumably
an old stellar population, a resonable range for the
microlens
mass is $0.1M_{\odot}  \leq M_{\rm \mu lens} \leq 1M_{\odot}$
(Alcock et al. 1997, Lewis \& Irwin 1995, Wyithe et al. 2000a).  
Using the length of the observing interval,
$t_{\rm obs}=126$  days, 
we  can  deduce  $v_{\rm upper}$, the upper limit
on the effective tranverse velocity in this lens system.

In calculating the effective transverse velocity of the lens in
the  lens plane  from  these numbers,  we  need to  use the  following
expression (Kayser et al. 1986):
\begin{equation}
\vec{V}=\frac{1}{1+z_s}~\vec{v_s}-\frac{1}{1+z_d}\frac{D_s}{D_d}
~\vec{v_d}+\frac{1}{1+z_d}\frac{D_{ds}}{D_d}~\vec{v_{obs}}
\label{eq5}
\end{equation}
where $\vec{V}$  is the effective  transverse velocity of  the system,
$\vec{v_s}$ the  velocity of the  source, $\vec{v_d}$ the  velocity of
the deflector  (lens), and $\vec{v_{\rm obs}}$  the velocity of  
the observer. The effective transverse motion of the lens includes the true
transverse velocity of the galaxy as a whole and an effective contribution due
to the stellar proper motions.
For the adopted concordance cosmology ($\Omega_o=0.3$,
$\Lambda_o=0.7$  and $H_0=65$~km~sec$^{-1}$~Mpc$^{-1}$)
and 
the  redshifts  of  the   
source  and  deflector, $z_s$ and $z_d$, respectively,
Equation~\ref{eq5} can be evaluated to get:
\begin{equation}
    {V}=0.37~{v_{s}} -10.55~{v_{d}}+10.18~{v_{\rm obs}}.
\label{eq6}
\end{equation}
Comparison of the Earth's motion   
relative to  the  microwave background 
(Lineweaver  et al.   1996) with the direction to the quasar Q2237+0305 
indicates  that these  vectors  are  almost  parallel, 
so  that the last  term on  the right  hand side of
Eq.~\ref{eq6} can be neglected: the transverse part is very small 
(in fact ~$v_{\rm obs}\simeq50$ km~sec$^{-1}$, and this value can in general be
known from the dipole term in the microwave background radiation and 
the direction towards the lens in any gravitational lens system).
Furthermore, assuming that the peculiar velocities of
the quasar and the lensing galaxy, ${v_s}$
and ${v_d}$,  are of the same order, 
the first term can be neglected as well, 
since its weight is only about $4\%
$ of the total. 
In this way, we just keep the expression
\begin{equation}
 {V} ~\simeq~10.55~{v_{d}}.
\end{equation}
An upper  limit for the effective transverse  velocity of the lens
in  the  lens  plane,  ${v_{d}}$,  
can now be  calculated  by just  setting
${V}=V_{\rm upper}$.
Here  we assumed 
$V_{\rm upper}$  to be the  90\% limit on the 
effective transverse velocity, i.e. it is based on the
90\% length $l_{\mathrm upper} =  0.13~r_E$,
as inferred  from the simulations.  
The resulting value for the 90\%-limit on the transverse
velocity of the lensing galaxy obtained in this way depends
on the assumed mass of the microlenses (and is independent
of the quasar size).
For $M_{\rm \mu lens}=0.1M_{\odot}$, we obtain:
$${v_{d, 0.1M_\odot, 90\%}}\leq 685~{\rm km/s}$$
for all the source sizes considered.
The limits for masses of 
$M_{\rm \mu lens}=1.0M_{\odot}$ is 
$${v_{d, 1.0M_\odot, 90\%}}\leq 2175~{\rm km/s}.$$ %
%
%


It is  even possible to place slightly stronger  
limits on ${v_{\rm bulk}}$. The reason is that 
the  actual effective lens velocity ${v_d}$ is a 
combination of the bulk velocity of  the galaxy as a whole
($    {v_{\rm bulk}}$)  and the velocity
dispersion of the microlenses  ($    {v_{\rm \mu lens}}$).
This latter effect was studied by 
Schramm et al. (1992), Wambsganss  \&  Kundic  (1993)
and  Kundi\'{c}  \& Wambsganss (1995), and it was found
that the two velocity contributions combined are 
producing the effective velocity in the following way:
\begin{equation}
    {v_d}=\sqrt{    {v_{\rm bulk}}^2+(a~{v_{\rm \mu lens}})^2}
\label{random}
\end{equation}
where $a$  represents the {\em effectiveness}  
of microlensing produced by the velocity dispersion
versus the bulk motion. As the velocity dispersion of compact
objects is more or less a random motion, the parameter $a$ is independent 
of the tranverse motion of the bulk.
The  value   of  this  `effectiveness  parameter'  is
$a~\approx~1.3$
(see  Wambsganss  \&  Kundic  1993, Kundi\'{c}  \&
Wambsganss 1995  for details).    
Since  the central velocity dispersion  of the lensing galaxy
in Q2237+0305 has been measured:  
$    {v_{\rm \mu  lens}}~\simeq~215$~km/s  (Foltz  et al. 1992),  
we can  use that and infer an even lower  value for 
the upper limit on the effective velocity  of the bulk motion 
(the velocity dispersion at the location of the images
could be slightly different, though):       
$$    {v_{\rm bulk, 0.1~M_\odot, 90\%}}~\leq~625~{\rm km/s} \ \ {\rm for}  \ \     
M_{\rm \mu lens}=0.1M_{\odot}$$
and
$$    {v_{\rm bulk, 1.1~M_\odot, 90\%}}~\leq~2157~{\rm km/s} 
	\ \ {\rm for}  \ \     M_{\rm \mu lens}=1.0M_{\odot}.$$
%
%
%
%

\section{Discussion}

Wyithe et al. (1999) presented the first determination of the
effective transverse velocity of the lens galaxy in Q2237 via 
microlensing. Here we compare this approach to ours. First, the
Wyithe et al. method uses a number of microlensing events, with
a base monitoring line of the order of 10 years or so. 
Our method --  based on
the amplitude of observed lightcurves -- can be applied to shorter monitoring
base lines (typically one order of magnitude lower) and is more restrictive in the 
absence of microlensing fluctuations. In principle, in the absence of microlensing 
events, both methods should yield similar results. Second, our statistics
is simple and straighforward: fluctuations higher than the observations are
ruled out in the simulations, no other assumptions are necessary. 
Third, our method is source size independent in the considered range. The Wyithe et al. 
results are slightly dependent on   the source size, although their largest upper limit 
is obtained for a point source (see their Fig.~12). On the other hand, Wyithe et al. 
constrain both mass and velocity simultanously, whereas in our case we assume we have 
some information on the mass range.	

More recently, Kochanek (2004) developed a general method for analyzing
microlensed quasar lightcurves, trying to simultaneously fit the effective
source velocity, the average stellar mass, the stellar mass function and the
size and structure of the quasar accretion disk. This ambitious task is
performed by a multidimensional $\chi^2$ statistic with an enormous
computational effort, but considers more realistic quasar models. Our
calculations for the effective transverse velocity are computationaly cheap and
easily applicable to other systems.

The result we adopt for the upper limit of the effective transverse velocity of
the lens galaxy (${v_{\rm bulk}}~\leq~630$~km/s, 90\% c.l., considering 
$M_{\rm \mu lens}=0.1M_{\odot}$) is slightly higher than 
that obtained by Wyithe et al. (${v_{\rm bulk}}~\leq~500$~km/s at a 95\% c.l.) 
but lower than that reported by Kochanek (2004) 
(${v_{\rm bulk}}~\leq~1800$~km/s, 68\% c.l, for 
$M_{\rm \mu lens}=0.1M_{\odot}$). Kochanek (2004) remarked that the 
significant variability of the quasar during the OGLE monitoring period 
(he analyzed the $V$-band OGLE 
data set) may be the cause of the relatively high velocities found 
in his work.

\section{Conclusions}

Estimating peculiar motions of galaxies is in general a difficult task. 
We have here derived upper limits to the transverse velocity of the lensing
galaxy in the quadruple quasar system Q2237$+$0305, 
using four months of monitoring data from the GLITP collaboration 
(Alcalde et al. 2002). Although we took the amplitude limits from all the 
lightcurves, the results are mainly constrained by components B and D, where 
no strong microlensing signals are present. 
The idea of the method is simple and straightforward:
if the galaxy is moving through the network of microcaustics and no or little 
microlensing is present in the observations, 
this defines a typical length of the low magnification regions
in the magnification patterns, which in turn can be 
converted into a physical velocity, when the length of the
observing interval is considered. 

This typical length is derived in a statistical sense from intensive 
numerical simulations using two different macro models
for the lens (which both produce the same results). 
The resulting value obtained for this upper limit on 
the transverse velocity   
of the lensing galaxy is 
$v_{\rm bulk, 0.1M_\odot, 90\%}\leq630$~km/s for lens masses of 
$M_{\rm \mu lens}=0.1M_{\odot}$ and 
$v_{\rm bulk, 1.0M_\odot, 90\%}\leq2160$~km/s for lens 
masses of $M_{\rm \mu lens}=1.0M_{\odot}$. 
Within the error 
estimation for this limit, the result is independent of
the quasar sizes considered 
(Gaussian width from 0.003 r$_E$ to 0.05 r$_E$, 
which corresponds to the range 
of $2\times10^4$ cm to $3\times10^5$ cm, 
in the case of $M_{\rm \mu lens}=0.1M_{\odot}$).
Future monitoring campaigns of this and other
multiply imaged quasars
can be used to provide more and stronger 
limits on the transverse velocities of lensing galaxies, in particular using
quiet periods in systems where microlensing has been previously detected.

\begin{acknowledgements}
The GLITP (Gravitational Lenses International Time Project) 
observations were  done on the Nordic Optical Telescope (NOT) \;-- from October
1st, 1999 to February 3rd, 2000 --,
which  is operated  on  the island  of  La Palma  jointly by  Denmark,
Finland, Iceland, Norway, and  Sweden, and is part of the Spanish Observatorio del
Roque  de  Los  Muchachos  of  the Instituto  de  Astrof\'{\i}sica  de
Canarias (IAC). We are grateful to the technical team of the telescope
for valuable  collaboration during  the observational work.  This work
was  supported  by  the Deutsche Forschungsgemeinschaft (DFG) grant 
WA~1047/6-1, the  P6/88  project of the  IAC,  Universidad  de 
Cantabria funds,  DGESIC (Spain) grant PB97-0220-C02  and the Spanish 
Department  for  Science  and  Technology  grants  AYA2000-2111-E  
and AYA2001-1647-C02.

\end{acknowledgements}
{}


\begin{thebibliography}{}

\bibitem[Alcalde et al. 2002]{Alcalde02}
Alcalde D., Mediavilla E., Moreau O. et al., 2002, ApJ, 572, 729

\bibitem[Alcock et al. 1997]{Alcock97}
Alcock C., Allsman R.A., Alves D. et al. (The MACHO Collaboration), 1997, ApJ,
491, L11

\bibitem[Chae et al. 1998]{Chae98}
Chae K-H., Turnshek D.A., Khersonsky V.K., 1998, ApJ, 495, 609

\bibitem[Dekel et al. 1990]{Dekel90}
Dekel A., Bertschinger E., Faber S.M., 1990, ApJ, 364, 349

\bibitem[Foltz et al. 1992]{Foltz92}
Foltz C.B., Hewett P.C., Webster R.L., Lewis G.F., 1992, ApJ, 386, L43

\bibitem[Goicoechea et al. 2002]{Goico02}
Goicoechea L.J., Alcalde D., Mediavilla E., Mu\~noz J.A., 2003, \aa, 397, 517

\bibitem[Gould 1995]{G95}
Gould A.,  1995, ApJ, 444, 556

\bibitem[Irwin et al. 1989]{I89} 
Irwin M.J., Webster R.L., Hewett P.C. et al., 1989, AJ, 98, 1989

\bibitem[Kayser et al. 1986]{Kayser86}
Kayser R., Refsdal S., Stabell R., 1986, \aa, 166, 36

\bibitem[Kochanek]{Ko04}
Kochanek C.S., 2004, ApJ, 605, 58


\bibitem[Kundic 1993]{KW93}
Kundi\'{c} T., Wambsganss J., 1993, ApJ, 404, 455

\bibitem[Lewis 1995]{LI95}
Lewis G.F., Irwin M.J., 1995, MNRAS, 276,103

\bibitem[Lineweaver et al. 1996]{LTSBL86}
Lineweaver C.H., Tenorio L., Smoot G.F. et al., 1996, ApJ, 470, 38

\bibitem[Ostensen et al. 1996]{Ost96} 
{\O}stensen R., Refsdal S., Stabell R. et al., 1996, \aa, 309, 59

\bibitem[Peebles 1993]{Peebles93}
Peebles P.J.E., 1993, "Principles of Physical Cosmology", Princeton University
Press.

\bibitem[Peebles et al. 2001]{Peebles01}
Peebles P.J.E., Phelps S.D., Shaya E.J., Tully R.B., 2001, ApJ, 554, 104

\bibitem[Roukema & Bajtlik 1999]{RB99}
Roukema B.F., Bajtlik S., 1999, MNRAS, 308, 309

\bibitem[Shalyapin et al. 2002]{Shalyapin02}
Shalyapin V.N., Goicoechea L.J., Alcalde D. et al., 2002, ApJ, 579, 127

\bibitem[Schmidt and Wambsganss 1998]{SW98}
Schmidt R., Wambsganss J., 1998, \aa, 335,379

\bibitem[Schmidt et al. 1998]{Schmidt98}
Schmidt R., Webster R.L., Lewis G.F., 1998, MNRAS, 295, 488 

\bibitem[Schneider et al. 1992]{SEF92}
Schneider P., Ehlers J., Falco E.E., 1992, "Gravitational Lenses", Springer
Verlag, Berlin

\bibitem[Schneider et al. 1988]{STG88}
Schneider D.P., Turner E.L., Gunn J.E. et al., 1988, AJ, 95, 1619 

\bibitem[Schramm et al. 1992]{Schramm92}
Schramm T., Kayser R., Chang K., et al., 1992, \aa, 268, 350

\bibitem[Wambsganss 1990]{W90}
Wambsganss J., 1990, PhD thesis (Munich University), also available
as report MPA 550 

\bibitem[Wambsganss 1994]{WP94}
Wambsganss J., Paczy\'{n}ski B., 1994, AJ, 108, 1156

\bibitem[Wambsganss 1995]{WK95}
Wambsganss J., Kundi\'{c} T., 1995, ApJ, 450, 19

\bibitem[Wambsganss 1999]{W99}
Wambsganss J., 1999, Journ. Comp. Appl. Math., 109, 353 

\bibitem[Witt et al. 1993]{WKR}
Witt H.J., Kayser R., Refsdal S., 1993, \aa, 268, 501

\bibitem[Witt and Mao 1994]{WM94}
Witt H.J., Mao S., 1994, ApJ, 429, 66

\bibitem[Wo{\'z}niak et al. 2000a]{Wozniak00a}
Wo\'{z}niak, P.R., Alard, C., Udalski, A., Szyma\'{n}ski, M., 
Kubiak, M., Pietrzy{\'n}ski, G., \& Zebru{\'n}, K. 2000a, ApJ, 529, 88

\bibitem[Wozniak et al. 2000]{Wozniak00b} 
Wo\'{z}niak P. R., Udalski A., Szyma\'{n}ski et al., 2000b, ApJ, 540, L65

\bibitem[Wyithe et al. 1999]{Wyithe99}
Wyithe J.S.B., Webster R.L., Turner E.L., 1999, MNRAS, 309, 261

\bibitem[Wyithe et al. 2000a]{Wyithe00a}
Wyithe J.S.B., Webster R.L., Turner E.L., 2000a, MNRAS, 315, 51

\bibitem[Wyithe et al. 2000b]{Wyithe00b}
Wyithe, J.S.B., Webster, R.L., Turner, E.L., 2000b, MNRAS, 318, 762

\bibitem[Yee 1988]{Yee88}
Yee H.K.C., 1988, AJ, 95, 1331

\bibitem[Yonehara 2001]{Yonehara01}
Yonehara, A., 2001, ApJ 548, L127

\end{thebibliography}
\end{document}